\documentclass[]{fairmeta}

\usepackage{amsmath,amssymb,amsfonts}
\usepackage{textcomp}
\usepackage{pifont}

\newcommand{\nve}[1]{\texttt{<#1>}}
\newcommand{\nvtablefont}{\small\sffamily}

\title{Beyond Words: Towards Effective Modeling of Non-Verbal Vocalizations in ASR}

\author{Gene-Ping Yang}
\author{Haibin Wu}
\author{Peng Su}
\author{Ruizhe Huang}
\author{Suwon Shon}
\author{Bach Do}
\author{Minxue Niu}
\author{Zhaoheng Ni}
\author{Shang-Wen Li}
\author{Florian Metze}
\author{Yossi Adi}
\author{Ming Sun}
\author{Yuzong Liu}

\affiliation{Meta}

\abstract{
Modern automatic speech recognition (ASR) systems excel at transcribing lexical content but often omit nonverbal vocalizations (NVs), such as laughter, breaths, coughs, and cries, that carry conversational and affective information.
Modeling NVs in ASR is challenging because NV annotations are sparse and highly long-tailed, with frequent categories such as breaths and laughter dominating rarer events such as cries and coughs.
We study three data-centric strategies for improving low-resource NV recognition: (1) a two-stage curriculum that first maps all NV events to a generic token and then fine-tunes on target categories; (2) inter-token transfer from high-resource events, such as laughter and breath, to rare events, such as crying; and (3) voice-conversion augmentation with class balancing.
Experiments show that shared acoustic structure across vocal events can be exploited to improve rare-category detection while preserving lexical ASR quality.
}

\date{\today}
\correspondence{Haibin Wu}
\metadata[Keywords]{ASR; Nonverbal Vocalization; Paralinguistic Event Detection}

\begin{document}

\maketitle

\section{Introduction}

Modern ASR systems have achieved strong lexical transcription performance, largely due to large-scale models trained on vast amounts of speech~\cite{radford2023whisper, baevski2020wav2vec, hsu2021hubert}.
Their outputs, however, often omit nonverbal vocalizations (NVs) such as laughter, sighs, coughs, cries, and breaths.
These events convey information about speaker state, affect, and conversational dynamics~\cite{schuller2013computational, schuller2013paralinguistics,zhang2024covomix,trouvain2012comparing}.
Ignoring them yields transcripts that are lexically accurate but incomplete for applications such as conversational AI, affective computing, and health-related speech analysis.

Directly integrating NV detection into end-to-end ASR raises two practical challenges.
First, many NV categories are extremely scarce: events such as \nve{cry} and \nve{sigh} are far less frequent than \nve{breath} and \nve{laugh}, making direct supervised training unreliable~\cite{matsuda2023detection,trouvain2012comparing,borisov2025nonverbaltss}.
This scarcity differs from rare-word recognition: rare words can reuse phones or subwords shared with other lexical items, whereas NV tags denote non-lexical vocal or oral events with no canonical phone sequence and variable duration~\cite{papadourakis2021phonetically,zhao2019shallow,lux2021meta}. 
Some NVs contain local acoustic cues that resemble speech sounds, such as airflow, closures, or releases, but they are not lexical consonants: they are event-level vocal or oral actions with variable duration and no canonical phone sequence~\cite{trouvain2014laughing}.
Second, NV data are heavily imbalanced.
In many data sets, \nve{breath} accounts for over 90\% of nonverbal samples, while \nve{cry} and \nve{cough} appear only rarely~\cite{trouvain2012comparing}.
Standard ASR objectives, which are dominated by frequent words and frequent NV tokens, can therefore suppress the learning signal for rare but important events~\cite{wang2019dynamic,chen2019rare}.

We address these challenges through three complementary strategies. 
First, we introduce a \textbf{two-stage curriculum learning framework} that separates general NV acoustic modeling from fine-grained rare-token specialization. 
In the first stage, all non-verbal events are mapped to a single generic label, e.g., \nve{NV}, so that the model can learn a broad acoustic foundation for non-speech vocal events from the full available data. 
In the second stage, the model is fine-tuned to distinguish specific low-resource target tokens, such as \nve{cough} and \nve{cry}, using only a small amount of category-specific supervision. 
This design reduces the need to learn rare NV categories from scratch and allows scarce labels to be used mainly for specialization.
Second, building on the observation that non-verbal vocalizations share a latent physiological and acoustic manifold~\cite{provine2000laughter, scherer2003vocal, trouvain2014laughing}, we leverage \textbf{inter-token knowledge transfer}. 
Many NV events, including breathing, laughter, sighing, and crying, are produced through related respiratory and laryngeal mechanisms, which suggests that high-resource NV categories may provide useful acoustic structure for low-resource ones~\cite{provine2000laughter, scherer2003vocal, trouvain2014laughing}. 
We use ``acoustic scaffold'' descriptively to refer to shared airflow, voicing characteristics, and temporal dynamics learned from high-resource NV categories.
By treating abundant events such as \nve{breath} and \nve{laugh} as this scaffold, we improve recognition of acoustically related but rare events like \nve{cry}.
Third, we address severe class imbalance using \textbf{class balancing} and \textbf{voice conversion (VC)} augmentation. 
Inspired by the metadata-balancing principles in MetaCLIP~\cite{xu2023demystifying,chuang2025meta}, we rebalance the training distribution so that rare NV categories contribute sufficient learning signal instead of being overwhelmed by dominant NV such as \nve{breath}. 
In addition, VC-based augmentation increases speaker diversity by generating speaker-varied versions of rare-class samples, which can reduce overfitting to the limited speaker identities present in scarce NV data. 
Together, these strategies provide a practical pathway for improving rare NV recognition while preserving the unified ASR formulation.

The main findings and contributions of this work are:
\begin{itemize}
    \item We demonstrate that a two-stage curriculum learning effectively improves the performance of learning rare NV categories (e.g. 100 cry samples in Table~\ref{tab:cry_improvement}).
    \item We provide empirical evidence for inter-token knowledge transfer, where using high-resource NVs like laughter and breath as an ``acoustic scaffold'' can more than double the recognition performance of rare events like crying, and that the choice of auxiliary category matters beyond its data volume (Table~\ref{tab:cry_mixture_improvement}).
    \item We show that class balancing combined with voice conversion is a practical solution for extreme class imbalance, enabling the recognition of categories that represent less than 2\% of training hours (Figure~\ref{fig:gigacms_bar_chart}).
    \item In a \textbf{system-level} comparison against Whisper-D\footnote{\url{https://huggingface.co/jordand/whisper-d-v1a}}, used here as an external reference rather than a controlled baseline, our model obtains higher F1 across all tested non-verbal categories (Table~\ref{tab:whisper_comparison}).

\end{itemize}

\section{Related Work}
\label{sec:related_work}

\subsection{Event-Aware Speech Recognition}

Conventional ASR targets lexical transcription, but spoken interaction also contains non-lexical events that carry affective, physiological, and conversational information.
Early end-to-end studies showed that verbal and nonverbal information can be modeled jointly by representing non-speech events as output symbols in the recognition sequence~\cite{inaguma2018end,shione2023automatic}.
More recent work has moved toward explicit event-aware recognition at larger scale.
NVSpeech introduces a pipeline that treats paralinguistic vocalizations as inline decodable tokens and uses the resulting recognizer to annotate a larger corpus for controllable speech synthesis~\cite{liao2025nvspeech}.
WESR further formalizes word-level event-speech recognition with a taxonomy of vocal events and a position-aware evaluation protocol that separates ASR errors from event detection and localization errors~\cite{wesr2026}.
These efforts establish that event tokens can be integrated into ASR outputs, but they mainly emphasize dataset construction, localization, or broad benchmark coverage.
Our work focuses on the complementary problem of training NV-aware ASR when target event categories are scarce and highly imbalanced.

\subsection{Nonverbal Vocalization Resources and Benchmarks}

Recent speech generation work has made nonverbal vocalizations, often abbreviated as NVs or NVVs, more visible by building datasets and benchmarks with explicit event inventories.
NonverbalTTS provides a public English corpus with text-aligned NV labels and emotion annotations for TTS~\cite{borisov2025nonverbaltss}, while NonVerbalSpeech-38K uses automatic annotation to scale natural-speech NV data across multiple categories~\cite{nonverbalspeech38k2025}.
Mandarin resources such as SMIIP-NV and MNV-17 similarly target expressive or performative nonverbal vocalizations with richer annotation schemes~\cite{wu2025smiipnv,mnv17_2025}.
On the evaluation side, NV-Bench proposes a functional taxonomy and metrics for expressive TTS with NVs~\cite{ni2026nvbench}, and NVV-SuperBench expands this direction with a 45-type bilingual taxonomy and multi-axis evaluation of controllability, placement, and perceptual salience~\cite{xue2026nvvsuperbench}.
These resources show that NV event coverage remains uneven across corpora: common events such as laughter and breathing are relatively well represented, whereas categories such as \nve{cry}, \nve{sigh}, \nve{cough}, \nve{smack}, and \nve{swallow} remain much less available. 
This imbalance motivates ASR training strategies that can transfer information from abundant NV categories to low-resource ones, rather than requiring a large labeled corpus for every target event.

\subsection{Learning With Long-Tailed Speech Events}

Long-tailed labels are difficult for sequence models because frequent words or events dominate the optimization signal, while rare targets receive limited supervision~\cite{wang2019dynamic,chen2019rare}.
Curriculum learning addresses related data-efficiency problems by starting from easier or more general targets before specializing to harder distinctions~\cite{bengio2009curriculum,karakasidis2024comparison}.
For low-resource speech, voice conversion and voice cloning can increase speaker diversity when real examples are scarce~\cite{bartelds2023making,liu2024seedvc}.
However, augmentation alone does not resolve label imbalance if rare events remain overwhelmed during training, so it must be paired with class balancing or other label-aware sampling strategies.
In parallel, phonetic and physiological studies suggest that laughter, crying, breathing, and sighing share respiratory and laryngeal structure~\cite{provine2000laughter,scherer2003vocal,trouvain2014laughing}, which creates an opportunity for transfer across NV categories.
Our experiments evaluate these ideas in an ASR setting by combining curriculum learning, auxiliary high-resource NV data, class balancing, and voice conversion augmentation.

\section{Method}

Our method targets the two main obstacles in NV-aware ASR: limited labeled data for rare events and strong class imbalance across NV categories.
We first describe the base ASR model and tokenization setup, then introduce the data-centric strategies used to improve rare-event recognition.

\subsection{Base ASR Model and NV Tokenization}
\label{sec:base_model}

Our ASR system is based on the RNN-Transducer (RNN-T) framework~\cite{graves2012sequence}, with an Emformer encoder~\cite{shi2021emformer}. 
The model predicts a single output sequence that contains both lexical tokens and non-verbal tokens. 
This allows NV detection to be integrated directly into ASR decoding, instead of relying on a separate event detector or a post-processing module.

Non-verbal events such as \nve{laugh}, \nve{cough}, and \nve{cry} are added as dedicated tokens in the SentencePiece vocabulary. 
During training, if an utterance contains a non-verbal event, the corresponding NV token is inserted into the transcript at the annotated position. 
The model is therefore trained to recognize words and NV events jointly. 
This formulation keeps the model architecture unchanged while making NV events visible to the sequence-level ASR objective.

In addition to the standard RNN-T loss, we apply a frame-level phone classification loss on the encoder.
For this auxiliary objective, all NV event frames are mapped to a shared spoken-noise phone label, \texttt{SPN}, following the convention used in hybrid HMM-based ASR systems~\cite{povey2011kaldi}. 
This auxiliary supervision encourages the encoder to learn frame-level representations of non-speech vocal activity. 
Because all NV categories share the same \texttt{SPN} label in this objective, rare categories can still benefit from the broader pool of non-verbal frames, even when their own token-level annotations are limited.

\subsection{Addressing Data Scarcity}
\label{sec:data_scarcity}

Data scarcity is one of the main obstacles for modeling rare NV categories. 
Some events, such as \nve{breath} and \nve{laugh}, occur frequently enough to support direct supervised training. 
Other events, such as \nve{cry}, are much rarer and are difficult to learn from token-specific examples alone. 
We address this problem with two complementary strategies: two-stage curriculum learning and inter-token knowledge transfer.

\subsubsection{Two-Stage Curriculum Training}

Inspired by curriculum learning~\cite{bengio2009curriculum}, we propose a two-stage training strategy that first learns a general representation for non-verbal sounds and then specializes this representation to individual NV tokens. 
The motivation is that coarse NV annotation is often easier to obtain than fine-grained labels for specific event types. 
For example, it may be easy to mark that an utterance contains a non-verbal sound, but much harder to collect enough accurately labeled examples of rare events such as \nve{cry}.

In Stage~1, all non-verbal events are mapped to a single generic token, \nve{NV}, regardless of their original category. 
The model is trained to distinguish speech from non-verbal vocal activity, without being required to separate different NV types. 
This stage allows the model to use all available NV data to learn a broad acoustic foundation for non-speech events. 
It also avoids the problem that rare tokens receive very little direct training signal when each NV category is modeled separately from the beginning.

In Stage~2, we restore the fine-grained NV labels and fine-tune the model with dedicated tokens such as \nve{cry}, \nve{laugh}, and \nve{breath}.
To transfer the knowledge learned in Stage~1, the embedding of the generic \nve{NV} token is copied to initialize the embeddings of \emph{all} individual NV tokens, and \nve{NV} itself is then removed from the output vocabulary.
This gives each NV token a useful starting point in the non-verbal acoustic space, instead of forcing the model to learn rare categories from scratch.

This design is especially useful for low-resource categories. 
For a rare event such as \nve{cry}, Stage~2 does not need to relearn what non-verbal vocal activity sounds like in general. 
It only needs to learn the acoustic cues that distinguish \nve{cry} from other NV events. 
In this work, we use \nve{cry} as the main case study because it is much less frequent than high-resource categories such as \nve{breath} and \nve{laugh}. 
The same strategy can also be applied to new rare NV categories, or to more fine-grained subcategories of existing events, such as distinguishing different types of laughter when only limited labeled data is available.

\subsubsection{Inter-token Knowledge Transfer}

Our second strategy is based on the observation that different non-verbal vocalizations often share physiological and acoustic structure. 
Events such as breathing, laughter, sighing, and crying are all related to respiratory and laryngeal activity~\cite{provine2000laughter, scherer2003vocal}. 
For example, laughter can be described as rhythmic glottal pulses over forced exhalation~\cite{trouvain2014laughing}. 
Although these events correspond to different labels, they may still occupy related regions of the acoustic space.

This motivates inter-token knowledge transfer. 
Instead of training a rare NV token only from its own limited examples, we use high-resource NV categories as auxiliary data. 
A category such as \nve{breath} can help the model learn basic airflow patterns and respiratory dynamics, while \nve{laugh} can provide additional information about voiced and rhythmic non-verbal vocalization. 
These learned acoustic patterns can then support recognition of a related but low-resource category such as \nve{cry}.

Prior work in speech synthesis also supports this intuition. 
For example, embeddings extracted from a laughter detection model have been shown to help with controllable crying synthesis~\cite{wu2024laugh}, suggesting that related vocal events can share useful representations. 
In our ASR setting, we test this idea by adding abundant \nve{laugh} and \nve{breath} examples as auxiliary training data for low-resource \nve{cry} recognition. 
This provides a practical way to improve rare NV modeling without requiring a large dedicated dataset for every new category.

\subsection{Addressing Data Imbalance}
\label{sec:data_imbalance}

\begin{figure}[t]
    \centering
    \includegraphics[width=\linewidth]{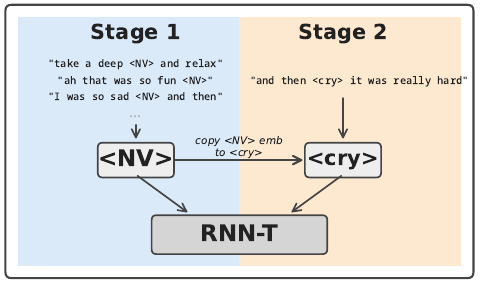}
    \caption{Two-stage curriculum learning pipeline. Stage~1 with \nve{NV} token, and Stage~2 with \nve{cry} token.
    }
    \label{fig:two_stage}
\end{figure}

\begin{figure}[t]
  \includegraphics[width=\linewidth]{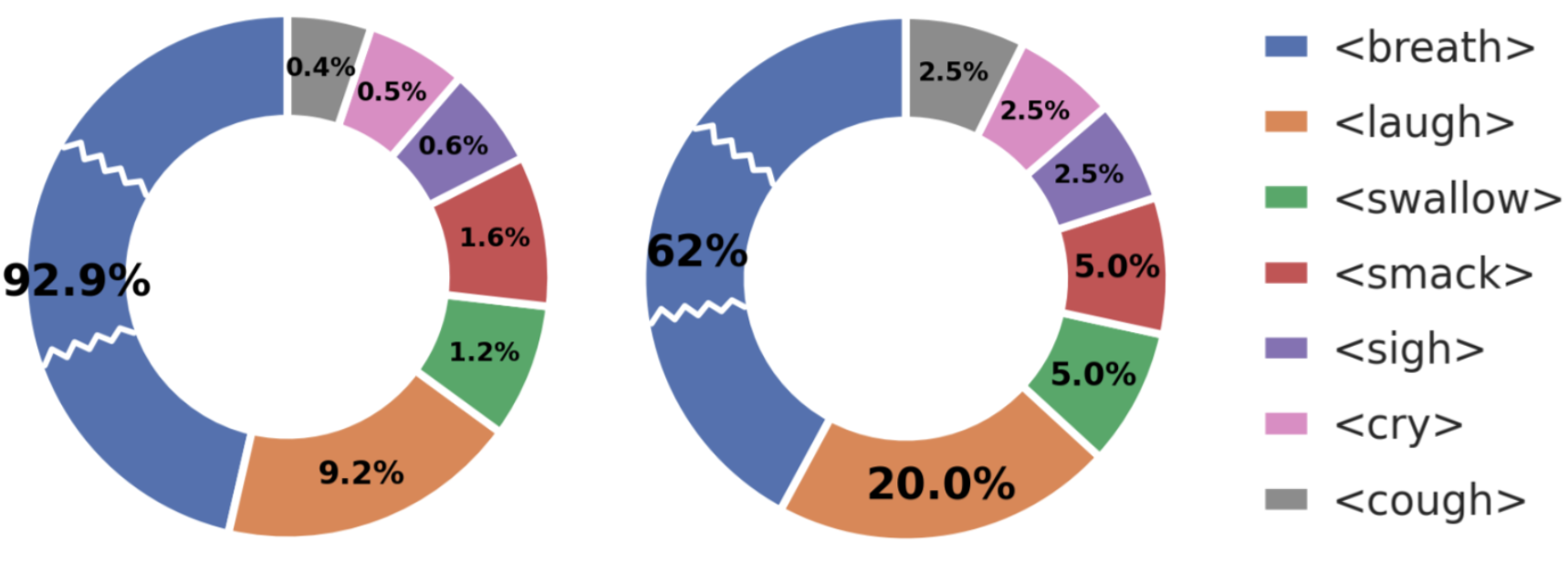}
  \makebox[0.35\linewidth]{(a) Original}
  \makebox[0.4\linewidth]{(b) Balanced}
  \caption{Sample distribution of NV labels before and after class balancing.}
  \label{fig:data_distribution}
\end{figure}

Beyond data scarcity, NV-ASR also suffers from severe class imbalance. 
In realistic speech data, frequent events such as \nve{breath} can dominate the NV distribution, while categories such as \nve{cry} or \nve{cough} may appear only rarely. 
If the model is trained directly on this skewed distribution, the learning signal is dominated by frequent categories, and rare events may be ignored even when they are important for downstream applications.

We use two strategies to reduce this imbalance. 
First, we apply class balancing by upsampling less frequent NV categories during training. 
This gives rare categories a stronger presence in the training batches and prevents high-frequency tokens from overwhelming the optimization process. 
As illustrated in Figure~\ref{fig:data_distribution}(b), this produces a more balanced NV distribution. 
To avoid excessive repetition of rare examples, we limit the upsampling ratio for rare categories to 2--5 times.
Class balancing is applied in both stages of the two-stage curriculum (Section~\ref{sec:data_scarcity}), so the generic \nve{NV} representation is not dominated by \nve{breath}.

Second, we use voice conversion augmentation to increase the diversity of rare NV samples. 
We apply Seed-VC~\cite{liu2024seedvc}, a zero-shot diffusion-based voice conversion system, to convert existing NV audio into the voices of 10 reference speakers. 
This creates up to $10\times$ more speaker-varied samples while preserving the acoustic characteristics of the original NV event. 
The goal is not to change the event label, but to reduce overfitting to the small number of speakers present in the rare-class data. 
By increasing speaker diversity, the model is encouraged to learn event-specific acoustic patterns that generalize better across speakers.

\section{Experimental setup}
\label{sec:experiments}

We use an in-house ASR model built on the RNN-Transducer framework, comprising an Emformer encoder \cite{graves2012sequence}, a predictor, and a joiner, with 200M parameters.
Our experiments use two in-house datasets, both annotated with seven NV sound categories: \nve{breath}, \nve{laugh}, \nve{swallow}, \nve{smack}, \nve{sigh}, \nve{cry}, and \nve{cough}.
The first is an 800-hour curated dataset with high-quality NV annotations, from which we reserve 1,700 utterances as our evaluation set.
The second is a 233-hour noisier dataset with lower annotation quality and an even more long-tailed NV distribution (as shown in Figure~\ref{fig:data_distribution}), which we use to study practical data augmentation and class balancing strategies.
In both datasets, \nve{breath} and \nve{laugh} are the most prevalent categories, reflecting their higher frequency in typical human vocalizations.
All experiments are conducted on English speech corpora.

All models are trained with Adam, weight decay $0.1$, and a tri-stage schedule peaking at $5\times10^{-6}$ for 20{,}000 updates.
Batches hold up to 100{,}000 frames per GPU across 32 GPUs.
The objective combines RNN-T loss with MWER loss \cite{prabhavalkar2018minimum} over 5-best hypotheses, weighted $0.2$ and $0.8$.

We evaluate NV prediction using Precision (P), Recall (R), and F1 per NV type.
The evaluation set contains 610 \nve{breath}, 408 \nve{laugh}, 131 \nve{swallow}, 206 \nve{smack}, 248 \nve{sigh}, 140 \nve{cry}, and 196 \nve{cough} instances; unlike the training distributions it is not long-tailed, so rare categories are well sampled.
Scoring is instance-based within a sentence, not presence-based: two reference \nve{laugh} tags with one predicted give one true positive and one false negative.
Matching ignores position, so localization is not measured.
We also report WER after removing NV tags from both hypothesis and reference, isolating the lexical ASR quality we aim to preserve.

\section{Experimental results}
We first conduct a system-level comparison between our best NV-ASR system and Whisper-D, where Whisper-D is used as an external reference system rather than a controlled training-data baseline; the comparison evaluates event-level behavior after canonical label mapping.
We then analyze the contribution of the proposed data-centric strategies, including two-stage curriculum learning, inter-token knowledge transfer, and the combination of class balancing with voice conversion (VC) augmentation. These experiments are designed to answer two questions: whether NV-aware training can improve non-verbal event recognition without degrading lexical ASR quality, and whether rare NV categories can be modeled effectively under severe data scarcity and class imbalance.

\subsection{Comparison with Whisper-D}

\begin{figure*}[!htp]
    \centering
    \includegraphics[width=\linewidth]{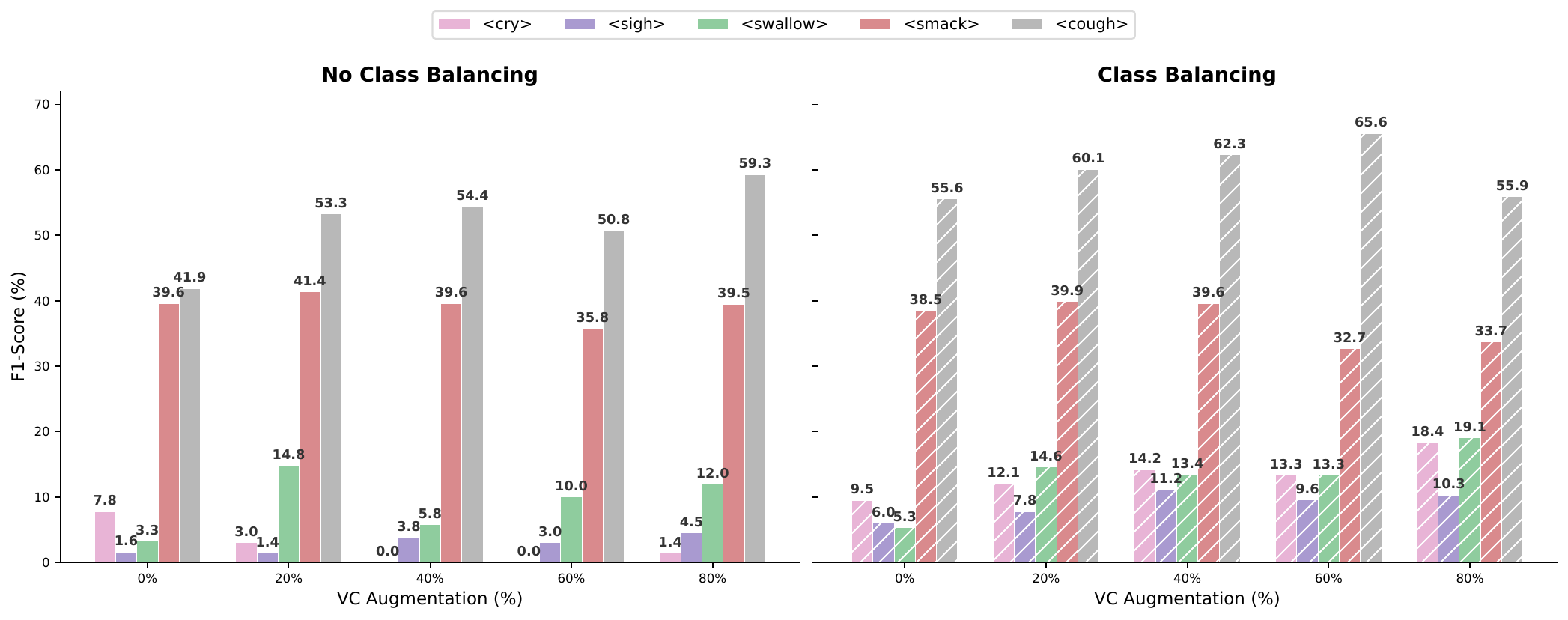}
    \caption{F1-scores of rare NV categories ($<$2\% of training hours) under different voice-conversion (VC) augmentation ratios. The left panel shows training without class balancing, and the right with class balancing. Class balancing makes VC augmentation more effective for long-tail NV detection.}
    \label{fig:gigacms_bar_chart}
\end{figure*}

\begin{table}[ht]
\centering
\nvtablefont
\caption{Comparison of NV detection performance between Whisper-D and our model.}
\label{tab:whisper_comparison}
\begin{tabular}{lcccccc}
\toprule
& \multicolumn{3}{c}{\textbf{Whisper-D}} & \multicolumn{3}{c}{\textbf{Ours}} \\
\cmidrule(lr){2-4} \cmidrule(lr){5-7}
\textbf{NV} & \textbf{P} & \textbf{R} & \textbf{F1} & \textbf{P} & \textbf{R} & \textbf{F1} \\
\midrule
breath  & 81.8	&  5.9 & 11.0 & 75.1 & 70.8 & \textbf{72.9} \\
laugh   & 68.9 & 65.7 & 67.3 & 79.1 & 88.2 & \textbf{83.4} \\
swallow & 100  &  0.8 &  1.5 & 95.4 & 78.6 & \textbf{86.2} \\
smack   & 66.7 &  2.9 &  5.6 & 72.5 & 56.3 & \textbf{63.4} \\
sigh    & 42.5 & 44.8 & 43.6 & 73.6 & 75.4 & \textbf{74.5} \\
cry     & 90.8 & 42.1 & 57.6 & 83.7 & 73.6 & \textbf{78.3} \\
cough   & 90.5 & 63.3 & 74.5 & 91.4 & 86.7 & \textbf{89.0} \\
\midrule
\textbf{WER (\%)}  & \multicolumn{3}{c}{2.40} & \multicolumn{3}{c}{\textbf{1.62}} \\
\bottomrule
\end{tabular}
\label{tab:nvasr}
\end{table}

Whisper-D was developed by fine-tuning the 1.55B-parameter Whisper-v2-large model~\cite{radford2023whisper} on approximately 22 hours of annotated speech from the Spotify Podcast Dataset~\cite{clifton-etal-2020-100000} containing human-labeled NV tags. Similar to Whisper-D, our system is initialized from a pretrained ASR checkpoint, but our model is substantially smaller, with approximately 200M parameters. For this comparison, we train our model on the 800-hour curated dataset, which provides high-quality NV annotations and consistent audio conditions for NV-ASR training. A key difference is that Whisper-D preserves heterogeneous surface forms from its training transcripts, such as \texttt{(sobs)}, \texttt{(crying)}, and \texttt{(Sobbing)} for crying-related events.
To ensure a fair event-level comparison, we post-map all Whisper-D outputs to the same canonical NV event inventory used by our model.

Table~\ref{tab:nvasr} shows that our model achieves higher F1-scores than Whisper-D for all evaluated NV categories. The strongest categories for our model are \nve{cough} (89.0), \nve{swallow} (86.2), and \nve{laugh} (83.4), while even lower-resource categories such as \nve{cry} and \nve{sigh} obtain strong F1-scores of 78.3 and 74.5, respectively. Whisper-D often shows high precision but very low recall for several categories, such as \nve{breath}, \nve{swallow}, and \nve{smack}, indicating that it misses many NV events after canonical mapping. In contrast, our model maintains a more balanced precision-recall trade-off across categories. Importantly, NV-aware training does not harm lexical ASR quality: our model reduces WER from 2.39\% for the seed checkpoint to 1.62\%, while Whisper-D reduces WER from 2.58\% to 2.40\% relative to its Whisper-v2-large baseline.

\subsection{Improving Low-Resource Detection via Two-stage Curriculum Learning}
\label{sec:curriculum}
\begin{table}[t!]
\centering
\nvtablefont
\caption{Effect of two-stage curriculum learning (CL) on cry detection across varying training set sizes.}
\label{tab:cry_improvement}
\begin{tabular}{cccccc}
\toprule
\textbf{Training Size} & \textbf{CL} & \textbf{P} & \textbf{R} & \textbf{F1} & \textbf{WER} \\
\midrule
100  & -          & 17.5 & 82.9 & 28.9 & 5.2 \\
100  & \checkmark & 22.4 & 91.4 & 36.0 & 5.2 \\
400  & -          & 19.8 & 84.3 & 32.1 & 3.3 \\
400  & \checkmark & 27.1 & 90.7 & 41.8 & 3.4 \\
4800 & -          & 26.5 & 90.7 & 41.0 & 2.5 \\
4800 & \checkmark & 30.6 & 94.3 & 46.2 & 2.2 \\
\bottomrule
\end{tabular}
\end{table}

To evaluate the effectiveness of two-stage curriculum learning (CL), we simulate different levels of \nve{cry} label scarcity by subsampling the 800-hour dataset at three scales: 100, 400, and 4,800 \nve{cry} samples. In Stage~1, all NV categories are mapped to a shared generic \nve{NV} token, allowing the model to learn a broad acoustic representation of non-verbal events from the full dataset. In Stage~2, the target \nve{cry} labels are restored for the selected subset, and the embedding of the generic \nve{NV} token is copied to initialize the \nve{cry} token embedding before fine-tuning. This design tests whether a coarse NV acoustic foundation can reduce the amount of fine-grained \nve{cry} supervision required for effective detection.

As shown in Table~\ref{tab:cry_improvement}, curriculum learning consistently improves \nve{cry} detection across all training sizes. With only 100 \nve{cry} samples, F1 increases from 28.9 to 36.0. With 400 samples, F1 improves from 32.1 to 41.8, giving an absolute gain of 9.7 points. Even at the 4,800-sample scale, curriculum learning further improves F1 from 41.0 to 46.2.
These gains indicate that the benefit of the two-stage strategy is not limited to the most extreme low-resource condition, although its impact is especially meaningful when labeled target data is scarce.
Because Table~\ref{tab:cry_improvement} is an ablation under restricted \nve{cry} supervision, it should not be directly compared with the fully resourced multi-category system in Table~\ref{tab:whisper_comparison}.

\subsection{Leveraging Auxiliary NV Data for Inter-token Knowledge Transfer}
\begin{table}[tp!]
\centering
\nvtablefont
\caption{Inter-token transfer from high-resource auxiliary NV data to low-resource \nve{cry} detection. All settings use only 400 labeled \nve{cry} utterances.}
\label{tab:cry_mixture_improvement}
\begin{tabular}{lrrrr}
\toprule
\textbf{Training Data} & \textbf{P} & \textbf{R} & \textbf{F1} & \textbf{WER} \\
\midrule
Cry only & 19.8 & 84.3 & 32.1 & 3.3 \\
+ Breath & 68.9 & 66.4 & 67.6 & 2.2 \\
+ Laugh  & 79.6 & 50.0 & 61.4 & 2.2 \\
+ Laugh + Breath & 88.8 & 56.4 & 69.0 & 1.9 \\
\bottomrule
\end{tabular}
\end{table}

We next study whether high-resource NV categories can improve recognition of a low-resource one. Using the same 400-sample \nve{cry} subset as in Section~\ref{sec:curriculum}, we augment the training data with 25,000 examples of \nve{breath}, \nve{laugh}, or both. The question is practical: instead of increasing the amount of \nve{cry} supervision directly, which is precisely what is hard to obtain, can abundant auxiliary NV events be used to make the few available \nve{cry} labels go further?
Table~\ref{tab:cry_mixture_improvement} shows that auxiliary NV data substantially improves \nve{cry} detection. Adding \nve{breath} data increases the \nve{cry} F1-score from 32.1 to \textbf{67.6}, an absolute gain of 35.5 points. Adding \nve{laugh} data also improves F1 to 61.4, with a particularly large precision gain from 19.8 to \textbf{79.6}. The best result is obtained by combining both \nve{laugh} and \nve{breath}, which raises the \nve{cry} F1-score to \textbf{69.0}. This is more than double the cry-only baseline and approaches the fully resourced \nve{cry} performance of 78.3 reported in Table~\ref{tab:whisper_comparison}.

These gains are not explained by data volume alone: \nve{breath} and \nve{laugh} add the same 25,000 utterances yet differ in F1 (67.6 vs.\ 61.4) and recall (66.4 vs.\ 50.0), so \emph{which} auxiliary category is added matters, not only how much.
We report this as a practical recipe rather than evidence for a specific transfer mechanism, since part of the precision gain is plausibly due to the added labels themselves: \nve{cry} is the only NV token in the cry-only setting.
Auxiliary data also improves overall ASR quality: WER drops from 3.3\% to \textbf{1.9\%} with both \nve{laugh} and \nve{breath}, approaching the 1.6\% of the full NV-ASR system in Table~\ref{tab:whisper_comparison}.

\subsection{Class Balancing and VC Augmentation}
\label{sec:balancing}

Complementing the above experiments on well-curated data, we next examine whether class balancing and augmentation can improve rare NV detection under realistic conditions.
To this end, we use our second, 233-hour in-house dataset, whose NV distribution exhibits an extreme long-tail pattern.
The training set exhibits an extremely long-tailed distribution: \nve{breath} accounts for over 90\% of NV-labeled segments, followed by \nve{laugh} with 17.7 hours (9.2\%). The remaining categories are severely underrepresented---\nve{smack} and \nve{swallow} each have roughly 4 hours, \nve{sigh} has 2.3 hours, \nve{cry} only 1.3 hours (0.5\%), and \nve{cough} a mere 0.2 hours (0.1\%).
To mitigate this imbalance, we applied two common remedies: \textbf{class balancing} to re-balance the training loss across categories, and \textbf{voice conversion (VC) augmentation} at varying data scale to synthetically increase data size and diversity.

Figure~\ref{fig:gigacms_bar_chart} summarizes the F1-scores of the five rare NV categories, each comprising less than 2\% of the training hours.
The leftmost bars of the left panel (0\% VC, no balancing) are a naive control: all seven NV tokens trained directly on the long-tailed data with none of the proposed strategies, while the remaining bars measure balancing and VC relative to it.
Our analysis reveals a critical synergy between the two remedies:
(1) At 0\% VC, data weighting (right panel) consistently outperforms the unweighted control (left panel) across all rare classes, with the exception of \nve{smack} which performs similarly.
Categories such as \nve{cry} and \nve{sigh}, which are near-zero in the control, show immediate modest gains once weighting is introduced.
(2) In the presence of data weighting, increasing VC augmentation from 0\% to 80\% provides a clear additional benefit, leading to improved detection for the long-tail categories.
Specifically, the \nve{cry} F1-score improves to 18.4\% at the 80\% VC level.
(3) From the left panel, without data weighting, the voice conversion augmentation provides no benefit to the lowest-resource categories, with the \nve{cry} F1-score even dropping to 0\% at 40--60\% VC.
This demonstrates that for extremely low-resource NV events, using VC augmentation only offers returns when combined with targeted class balancing.
Together, these findings reinforce that while data-centric remedies are helpful for skewed distributions, the targeted use of in-domain auxiliary data remains the most effective pathway for high-quality detection.

\section{Discussion}
\label{sec:discussion}

The experiments point to two practical lessons for scaling NV-aware ASR.
First, rare NV categories should not be treated as isolated labels.
The large gains from adding \nve{laugh} and \nve{breath} data are consistent with the model benefiting from shared respiratory and vocal-tract patterns before making a fine-grained decision about \nve{cry}.
This is consistent with the curriculum results: when the model first learns a generic NV region, it can use a small number of labeled target examples more effectively.
For emerging NV categories, this suggests a practical annotation strategy: collect a small set of high-quality target labels, but train them together with larger pools of related nonverbal events.

Second, augmentation and balancing address different failure modes.
Voice conversion increases speaker diversity, but it does not by itself change how much the rare class contributes to the training objective.
This explains why VC alone can fail under extreme imbalance, while VC combined with class balancing improves long-tail detection.
In other words, synthetic diversity is useful only when the optimization process is forced to pay attention to the rare categories.
This observation is important for NV-ASR because many rare events are not merely data-poor; they are also surrounded by much larger amounts of speech and frequent NV tokens.

There are also limitations.
Our evaluation focuses on a fixed set of seven NV categories, and the main low-resource case study is \nve{cry}.
The same principles should extend to other emerging labels, but the amount of useful transfer may depend on acoustic similarity between source and target events.
In addition, our metrics count NV tag instances within a sentence and report lexical WER after removing NV tags; because matching is by count rather than position, they do not measure fine temporal localization of each event.
Future work should combine rare-event training strategies with position-aware evaluation and broader NV taxonomies, especially for subtle oral cues and sustained affective vocalizations that remain difficult for current systems.

\section{Conclusion}
\label{sec:conclusion}

This paper studies practical, data-centric strategies for extending ASR systems to low-resource NV categories.
Across curated and long-tailed datasets, we find that rare-event performance depends not only on the number of target examples, but also on how the model is exposed to related non-speech acoustics.
Two-stage curriculum learning improves data efficiency by first learning a generic NV representation, and auxiliary \nve{laugh} and \nve{breath} data raise \nve{cry} F1 to 69.0\% with only 400 target samples.
Class balancing and voice conversion provide additional gains under severe distribution skew, especially when used together.
These findings suggest a scalable path for adding emerging nonverbal categories to ASR without collecting a large dedicated corpus for every event type.



\clearpage
\bibliographystyle{assets/plainnat}
\bibliography{paper}

@book{provine2000laughter,
  title  = {Laughter: A scientific investigation},
  author = {Provine, Robert R},
  year   = {2000},
  publisher = {Penguin}
}

@article{scherer2003vocal,
  title  = {Vocal communication of emotion: A review of research paradigms},
  author = {Scherer, Klaus R},
  year   = {2003},
  journal = {Speech communication},
  publisher = {Elsevier},
  volume = {40},
  number = {1-2},
  pages  = {227--256}
}

@inproceedings{bengio2009curriculum,
  title  = {Curriculum learning},
  author = {Bengio, Yoshua and Louradour, J{\'e}r{\^o}me and Collobert, Ronan and Weston, Jason},
  year   = {2009},
  booktitle = {Proceedings of the 26th Annual International Conference on Machine Learning},
  pages  = {41--48}
}

@inproceedings{povey2011kaldi,
  title  = {The Kaldi speech recognition toolkit},
  author = {Povey, Daniel and Ghoshal, Arnab and Boulianne, Gilles and Burget, Luk{\'a}s and Glembek, Ond{\v{r}}ej and Goel, Nagendra and Hannemann, Mirko and Motl{\'i}{\v{c}}ek, Petr and Qian, Yanmin and Schwarz, Petr and others},
  year   = {2011},
  booktitle = {IEEE 2011 Workshop on Automatic Speech Recognition and Understanding},
  organization = {IEEE Signal Processing Society}
}

@inproceedings{graves2012sequence,
  title  = {Sequence transduction with recurrent neural networks},
  author = {Graves, Alex},
  year   = {2012},
  booktitle = {ICML 2012 Workshop on Representation Learning}
}

@inproceedings{trouvain2012comparing,
  title  = {Comparing non-verbal vocalisations in conversational speech corpora},
  author = {Trouvain, J{\"u}rgen and Truong, Khiet P.},
  year   = {2012},
  booktitle = {Proceedings of the 4th International Workshop on Corpora for Research on Emotion Sentiment and Social Signals (ES3)},
  pages  = {36--39}
}

@book{schuller2013computational,
  title  = {Computational Paralinguistics: Emotion, Affect and Personality in Speech and Language Processing},
  author = {Schuller, Bj{\"o}rn and Batliner, Anton},
  year   = {2013},
  publisher = {Wiley}
}

@article{schuller2013paralinguistics,
  title  = {Paralinguistics in speech and language---State-of-the-art and the challenge},
  author = {Schuller, Bj{\"o}rn and Steidl, Stefan and Batliner, Anton and Burkhardt, Felix and Devillers, Laurence and M{\"u}ller, Christian and Narayanan, Shrikanth},
  year   = {2013},
  journal = {Computer Speech \& Language},
  publisher = {Elsevier},
  volume = {27},
  number = {1},
  pages  = {4--39}
}

@inproceedings{trouvain2014laughing,
  title  = {Laughing, breathing, clicking-The prosody of nonverbal vocalizations},
  author = {Trouvain, J{\"u}rgen},
  year   = {2014},
  booktitle = {Proceedings of the 7th international conference on Speech Prosody},
  pages  = {598--602}
}

@inproceedings{prabhavalkar2018minimum,
  title  = {Minimum word error rate training for attention-based sequence-to-sequence models},
  author = {Prabhavalkar, Rohit and Sainath, Tara N and Wu, Yonghui and Nguyen, Patrick and Chen, Zhifeng and Chiu, Chung-Cheng and Kannan, Anjuli},
  year   = {2018},
  booktitle = {2018 IEEE International Conference on Acoustics, Speech and Signal Processing (ICASSP)},
  pages  = {4839--4843},
  organization = {IEEE}
}

@inproceedings{inaguma2018end,
  title  = {An end-to-end approach to joint social signal detection and automatic speech recognition},
  author = {Inaguma, Hirofumi and Mimura, Masato and Inoue, Koji and Yoshii, Kazuyoshi and Kawahara, Tatsuya},
  year   = {2018},
  booktitle = {2018 IEEE International Conference on Acoustics, Speech and Signal Processing (ICASSP)},
  pages  = {5199--5203},
  organization = {IEEE}
}

@inproceedings{wang2019dynamic,
  title  = {Dynamic curriculum learning for imbalanced data classification},
  author = {Wang, Yiru and Gan, Wei and Yang, Jie and Wu, Wei and Yan, Jun},
  year   = {2019},
  booktitle = {Proceedings of the IEEE/CVF international conference on computer vision},
  pages  = {5017--5026}
}

@inproceedings{chen2019rare,
  title  = {Rare sound event detection using deep learning and data augmentation},
  author = {Chen, Xi and Kumar, Anurag and Narayanan, Shrikanth},
  year   = {2019},
  booktitle = {Proc. Interspeech 2019},
  pages  = {1218--1222}
}

@inproceedings{clifton-etal-2020-100000,
  title  = {100,000 Podcasts: A Spoken {E}nglish Document Corpus},
  author = {Clifton, Ann  and Reddy, Sravana  and Yu, Yongze  and Pappu, Aasish  and Rezapour, Rezvaneh  and Bonab, Hamed  and Eskevich, Maria  and Jones, Gareth  and Karlgren, Jussi  and Carterette, Ben  and Jones, Rosie},
  year   = {2020},
  month  = dec,
  booktitle = {Proceedings of the 28th International Conference on Computational Linguistics},
  publisher = {International Committee on Computational Linguistics},
  address = {Barcelona, Spain (Online)},
  pages  = {5903--5917},
  doi    = {10.18653/v1/2020.coling-main.519},
  editor = {Scott, Donia  and Bel, Nuria  and Zong, Chengqing}
}

@inproceedings{baevski2020wav2vec,
  title  = {wav2vec 2.0: A framework for self-supervised learning of speech representations},
  author = {Baevski, Alexei and Zhou, Henry and Mohamed, Abdel-rahman and Auli, Michael},
  year   = {2020},
  booktitle = {Advances in Neural Information Processing Systems},
  volume = {33},
  pages  = {12449--12460}
}

@article{hsu2021hubert,
  title  = {HuBERT: Self-supervised speech representation learning by masked prediction of hidden units},
  author = {Hsu, Wei-Ning and Bolte, Benjamin and Tsai, Yao-Hung Hubert and Lakhotia, Kushal and Salakhutdinov, Ruslan and Mohamed, Abdelrahman},
  year   = {2021},
  journal = {IEEE/ACM Transactions on Audio, Speech, and Language Processing},
  publisher = {IEEE},
  volume = {29},
  pages  = {3451--3460}
}

@inproceedings{shi2021emformer,
  title  = {Emformer: Efficient memory transformer based acoustic model for low latency streaming speech recognition},
  author = {Shi, Yangyang and Wang, Yongqiang and Wu, Chunyang and Yeh, Ching-Feng and Chan, Julian and Zhang, Frank and Le, Duc and Seltzer, Mike},
  year   = {2021},
  booktitle = {ICASSP 2021-2021 IEEE International Conference on Acoustics, Speech and Signal Processing (ICASSP)},
  pages  = {6783--6787},
  organization = {IEEE}
}

@inproceedings{radford2023whisper,
  title  = {Robust Speech Recognition via Large-Scale Weak Supervision},
  author = {Radford, Alec and Kim, Jong Wook and Xu, Tao and Brockman, Greg and McLeavey, Christine and Sutskever, Ilya},
  year   = {2023},
  booktitle = {International Conference on Machine Learning},
  pages  = {28492--28518},
  organization = {PMLR}
}

@inproceedings{matsuda2023detection,
  title  = {Detection of laughter and screaming using the attention and CTC models},
  author = {Matsuda, Tokuto and Arimoto, Yoshiko},
  year   = {2023},
  booktitle = {Proc. Interspeech 2023},
  pages  = {2268--2272}
}

@article{shione2023automatic,
  title  = {Automatic Speech Recognition Using Linguistic and Verbal/Nonverbal Information},
  author = {Shione, Nagisa and Kawahara, Tatsuya},
  year   = {2023},
  journal = {IEEE Access},
  publisher = {IEEE},
  volume = {11},
  pages  = {138985--138995}
}

@inproceedings{bartelds2023making,
  title  = {Making More of Little Data: Improving Low-Resource Automatic Speech Recognition Using Data Augmentation},
  author = {Bartelds, Martijn  and San, Nay  and McDonnell, Bradley  and Jurafsky, Dan  and Wieling, Martijn},
  year   = {2023},
  month  = jul,
  booktitle = {Proceedings of the 61st Annual Meeting of the Association for Computational Linguistics (Volume 1: Long Papers)},
  publisher = {Association for Computational Linguistics},
  address = {Toronto, Canada},
  pages  = {715--729},
  editor = {Rogers, Anna  and Boyd-Graber, Jordan  and Okazaki, Naoaki}
}

@article{karakasidis2024comparison,
  title  = {Comparison and analysis of new curriculum criteria for end-to-end ASR},
  author = {Karakasidis, Georgios and Kurimo, Mikko and Bell, Peter and Gr{\'o}sz, Tam{\'a}s},
  year   = {2024},
  journal = {Speech Communication},
  publisher = {Elsevier},
  volume = {163},
  pages  = {103113}
}

@article{zhang2024covomix,
  title  = {Covomix: Advancing zero-shot speech generation for human-like multi-talker conversations},
  author = {Zhang, Leying and Qian, Yao and Zhou, Long and Liu, Shujie and Wang, Dongmei and Wang, Xiaofei and Yousefi, Midia and Qian, Yanmin and Li, Jinyu and He, Lei and others},
  year   = {2024},
  journal = {Advances in Neural Information Processing Systems},
  volume = {37},
  pages  = {100291--100317}
}

@article{liu2024seedvc,
  title  = {Zero-shot Voice Conversion with Diffusion Transformers},
  author = {Liu, Songting},
  year   = {2024},
  journal = {arXiv preprint arXiv:2411.09943}
}

@inproceedings{xu2023demystifying,
  title  = {Demystifying {CLIP} Data},
  author = {Hu Xu and Saining Xie and Xiaoqing Tan and Po-Yao Huang and Russell Howes and Vasu Sharma and Shang-Wen Li and Gargi Ghosh and Luke Zettlemoyer and Christoph Feichtenhofer},
  year   = {2024},
  booktitle = {The Twelfth International Conference on Learning Representations}
}

@inproceedings{wu2024laugh,
  title  = {Laugh now cry later: Controlling time-varying emotional states of flow-matching-based zero-shot text-to-speech},
  author = {Wu, Haibin and Wang, Xiaofei and Eskimez, Sefik Emre and Thakker, Manthan and Tompkins, Daniel and Tsai, Chung-Hsien and Li, Canrun and Xiao, Zhen and Zhao, Sheng and Li, Jinyu and Kanda, Naoyuki},
  year   = {2024},
  booktitle = {2024 IEEE Spoken Language Technology Workshop (SLT)},
  pages  = {690--697},
  organization = {IEEE}
}

@inproceedings{chuang2025meta,
  title  = {Meta clip 2: A worldwide scaling recipe},
  author = {Chuang, Yung-Sung and Li, Yang and Wang, Dong and Yeh, Ching-Feng and Lyu, Kehan and Raghavendra, Ramya and Glass, James and Huang, Lifei and Weston, Jason and Zettlemoyer, Luke and Chen, Xinlei and Liu, Zhuang and Xie, Saining and Yih, Scott and Li, Shang-Wen and Xu, Hu},
  year   = {2025},
  booktitle = {Advances in Neural Information Processing Systems}
}

@article{liao2025nvspeech,
  title  = {{NVSpeech}: An Integrated and Scalable Pipeline for Human-Like Speech Modeling with Paralinguistic Vocalizations},
  author = {Liao, Huan and Ni, Qinke and Wang, Yuancheng and Lu, Yiheng and Zhan, Haoyue and Xie, Pengyuan and Zhang, Qiang and Wu, Zhizheng},
  year   = {2025},
  journal = {arXiv preprint arXiv:2508.04195}
}

@article{wesr2026,
  title  = {{WESR}: Scaling and Evaluating Word-Level Event-Speech Recognition},
  author = {Yang, Chenchen and Huang, Kexin and Fan, Liwei and Tu, Qian and Jiang, Botian and Zhang, Dong and Yin, Linqi and Li, Shimin and Fei, Zhaoye and Cheng, Qinyuan and Qiu, Xipeng},
  year   = {2026},
  journal = {arXiv preprint arXiv:2601.04508}
}

@inproceedings{borisov2025nonverbaltss,
  title  = {{NonverbalTTS}: A Public {English} Corpus of Text-Aligned Nonverbal Vocalizations with Emotion Annotations for Text-to-Speech},
  author = {Borisov, Maksim and Spirin, Egor and Diatlova, Daria},
  year   = {2025},
  booktitle = {Proc. SSW 2025},
  pages  = {104--109}
}

@article{nonverbalspeech38k2025,
  title  = {A Scalable Pipeline for Enabling Non-Verbal Speech Generation and Understanding},
  author = {Ye, Runchuan and Zhou, Yixuan and Yu, Renjie and Lin, Zijian and Li, Kehan and Li, Xiang and Liu, Xin and Zeng, Guoyang and Wu, Zhiyong},
  year   = {2025},
  journal = {arXiv preprint arXiv:2508.05385}
}

@inproceedings{wu2025smiipnv,
  title  = {{SMIIP-NV}: A Multi-Annotation Non-Verbal Expressive Speech Corpus in {Mandarin} for LLM-Based Speech Synthesis},
  author = {Wu, Zhuojun and Liu, Dong and Liu, Juan and Wang, Yechen and Li, Linxi and Jin, Liwei and Bu, Hui and Zhang, Pengyuan and Li, Ming},
  year   = {2025},
  booktitle = {Proceedings of the 33rd ACM International Conference on Multimedia},
  pages  = {12564--12570}
}

@inproceedings{mnv17_2025,
  title  = {{MNV-17}: A High-Quality Performative {Mandarin} Dataset for Nonverbal Vocalization Recognition in Speech},
  author = {Mai, Jialong and Ji, Jinxin and Xing, Xiaofen and Yang, Chen and Chen, Weidong and Xing, Jingyuan and Xu, Xiangmin},
  year   = {2026},
  booktitle = {{IEEE International Conference on Acoustics, Speech and Signal Processing}},
  pages  = {18312--18316},
  organization = {IEEE}
}

@article{ni2026nvbench,
  title  = {{NV-Bench}: Benchmark of Nonverbal Vocalization Synthesis for Expressive Text-to-Speech Generation},
  author = {Ni, Qinke and Liao, Huan and Chen, Dekun and Wang, Yuxiang and Wu, Zhizheng},
  year   = {2026},
  journal = {arXiv preprint arXiv:2603.15352}
}

@article{xue2026nvvsuperbench,
  title  = {{NVV-SuperBench}: Beyond Words, Beyond Quality-benchmarking Nonverbal Vocalizations in Speech Generation},
  author = {Xue, Liumeng and Bian, Weizhen and Pan, Jiahao and Wu, Wenxuan and Ren, Yilin and Kang, Boyi and Hu, Jingbin and Ma, Ziyang and Wang, Shuai and Qian, Xinyuan and Lee, Hung-yi and Guo, Yike},
  year   = {2026},
  journal = {arXiv preprint arXiv:2604.16211}
}

@inproceedings{papadourakis2021phonetically,
  title={Phonetically Induced Subwords for End-to-End Speech Recognition},
  author={Papadourakis, Vasileios and M{\"u}ller, Markus and Liu, Jing and Mouchtaris, Athanasios and Omologo, Maurizio},
  booktitle={Proc. Interspeech 2021},
  pages={2576--2580},
  year={2021},
  organization={ISCA}
}

@inproceedings{zhao2019shallow,
  title={Shallow-Fusion End-to-End Contextual Biasing},
  author={Zhao, Ding and Sainath, Tara N and Rybach, David and Rondon, Pat and Bhatia, Deepti and Li, Bo and Pang, Ruoming},
  booktitle={Proc. Interspeech 2019},
  pages={1418--1422},
  year={2019},
  organization={ISCA}
}

@inproceedings{lux2021meta,
  title={Meta-Learning for Improving Rare Word Recognition in End-to-End ASR},
  author={Lux, Florian and Vu, Ngoc Thang},
  booktitle={ICASSP 2021-2021 IEEE International Conference on Acoustics, Speech and Signal Processing (ICASSP)},
  pages={5999--6003},
  year={2021},
  organization={IEEE}
}

\end{document}